\documentstyle[psfig,12pt,axodraw]{article}
\begin{document}
\baselineskip 18pt
\def\today{\ifcase\month\or
 January\or February\or March\or April\or May\or June\or
 July\or August\or September\or October\or November\or December\fi
 \space\number\day, \number\year}
\def\thebibliography#1{\section*{References\markboth
 {References}{References}}\list
 {[\arabic{enumi}]}{\settowidth\labelwidth{[#1]}
 \leftmargin\labelwidth
 \advance\leftmargin\labelsep
 \usecounter{enumi}}
 \def\newblock{\hskip .11em plus .33em minus .07em}
 \sloppy
 \sfcode`\.=1000\relax}
\let\endthebibliography=\endlist
\begin{titlepage}
\hspace*{10.0cm}ICRR-Report-389-97-12 
   
\hspace*{10.0cm}OCHA-PP-96
\  \
\vskip 0.5 true cm 
\begin{center}
{\large {\bf Implications of baryon asymmetry for }}  \\
{\large {\bf the electric dipole moment of the neutron }}
\vskip 2.0 true cm
\renewcommand{\thefootnote}
{\fnsymbol{footnote}}
Mayumi Aoki \footnote{Research Fellow of the Japan Society 
for the Promotion of Science.},  
Akio Sugamoto
\\
\vskip 0.5 true cm 
{\it Department of Physics {\rm and} 
Graduate School of Humanities and Sciences } \\
{\it Ochanomizu University}  \\
{\it Otsuka 2-1-1, Bunkyo-ku, Tokyo 112, Japan}  \\
\vspace{1cm}
Noriyuki Oshimo
\\
\vskip 0.5 true cm
{\it Institute for Cosmic Ray Research} \\
{\it University of Tokyo} \\
{\it Midori-cho 3-2-1, Tanashi, Tokyo 188, Japan}  \\
\end{center}

\vskip 1.0 true cm

\centerline{\bf Abstract}
\medskip
     We study baryogenesis at the 
electroweak phase transition of the universe within the framework of  
the supersymmetric standard model (SSM) based on $N=1$ supergravity.  
This model contains a new source of $CP$ violation in  
the mass-squared matrices for squarks, 
which could enable $t$ squarks to mediate 
the charge transport mechanism for generating baryon asymmetry.  
The same $CP$-violating source also  
induces the electric dipole moment (EDM) of the neutron 
at the one-loop level.  
If the new $CP$-violating phase is not suppressed, it is shown, 
the $t$-squark transport can lead to  
baryon asymmetry consistent   
with its observed value within reasonable ranges of 
SSM parameters.  For these parameter ranges the magnitude of 
the neutron EDM is predicted to be not much smaller than 
its present experimental upper bound.  

\end{titlepage}

\def\gsim{{\mathop >\limits_\sim}}
\def\lsim{{\mathop <\limits_\sim}}
\def\r2{\sqrt 2}
\def\sw2{\sin^2\theta_W}
\def\v#1{v_#1}
\def\tb{\tan\beta}
\def\c2b{\cos 2\beta}
\def\sq{\tilde q}
\def\st{\tilde t}
\def\m#1{{\tilde m}_#1}
\def\mH{m_H}
\def\mg{{\tilde m}_g}
\def\M{\tilde M}
\def\mgr{m_{3/2}}
\def\dW{\delta_W}
\def\vW2{v_W^2}
\def\muB{\mu_B}

     The invariance for $CP$ transformation is not respected 
in nature.  Up to now $CP$ violation has only been observed 
in the $K^0$-$\bar K^0$ system, 
which can be well described by the Kobayashi-Maskawa (KM) 
mechanism of the standard model (SM).  
On the other hand, 
baryon asymmetry of our universe could also be an 
outcome of $CP$ violation \cite{ewbrev}, which was shown not to be explained 
by the SM.  Some extension for the SM seems to be 
necessary for baryogenesis.  In fact,  
the SM can only generate baryon asymmetry much smaller than 
its observed value.  
However, it may nevertheless be understood 
within the framework of physics at the electroweak scale, 
if there exists some new source of $CP$ violation 
other than  the KM mechanism.  

     The supersymmetric standard model (SSM) based on 
$N=1$ supergravity \cite{ssmrev}
is one of the most plausible extensions of the SM 
at the electroweak scale, which has new 
$CP$-violating phases in addition to the standard KM phase.   
These new sources of $CP$ violation do not affect much 
the $K^0$-$\bar K^0$ or $B^0$-$\bar B^0$ system, 
while they could induce baryon asymmetry.  
In particular, under the condition that the electroweak 
phase transition of the universe is strongly first order,  
an enough amount of asymmetry may be generated 
through the charge transport mechanism mediated by 
charginos or $t$ squarks \cite{nelson1,nelson2,aoki}.  
On the other hand, the same $CP$-violating phases 
give contributions to the electric dipole moments (EDMs) of 
the neutron and the electron at the one-loop level.  
If the baryon asymmetry is really attributed to the new phases, 
the EDMs will be predicted to have sizable values.  

     In this letter we study the possibility of baryogenesis 
in the SSM and its implications for the EDMs of the neutron 
and the electron.  
It is shown that the charge transport mechanism mediated by 
$t$ squarks can explain the baryon asymmetry, 
provided that the relevant $CP$-violating phase is not 
much suppressed.  The masses of supersymmetric 
particles which allow the baryogenesis are mostly 
in the ranges of 100$-$1000 GeV.    
In these parameter ranges the EDMs of the neutron and the electron 
have magnitudes not much smaller than their  
present experimental upper bounds, 
which would be examined experimentally in the near future.   

     We first discuss new sources of $CP$ violation in the SSM.    
For the physical complex 
parameters intrinsic in this model, without loss of generality, we can take
a Higgsino mass parameter $\mH$ for the bilinear term 
of the Higgs superfields in  
superpotential and several dimensionless coupling 
constants $A_f$ for the trilinear terms of the 
scalar fields which break supersymmetry softly.   
We express these parameters as 
\begin{eqnarray}
\mH &=& |\mH|\exp(i\theta), \nonumber \\
A_f &=& A = |A|\exp(i\alpha).   
\label{1}
\end{eqnarray}
Since $A_f$'s are considered to have the same value of 
order unity at the grand unification scale, 
their differences at the electroweak scale are 
small and thus can be neglected.    

     The new sources of $CP$ violation give 
contributions to the EDMs of the neutron and the electron 
through one-loop diagrams in which the charginos, neutralinos, 
or gluinos are exchanged together with the squarks or sleptons.  
If the $CP$-violating phase $\theta$ is of order unity, 
the neutron and electron EDMs receive large contributions 
from the chargino diagrams.  From the experimental constraints, 
the masses of the squarks and sleptons are then predicted 
to be larger than 1 TeV, while the 
charginos and neutralinos could have
masses of order 100 GeV \cite{edm}. 
In this case, it has been shown \cite{aoki},  
the charginos can mediate the charge transport 
mechanism and generate an enough amount of baryon asymmetry.  
However, some extension of the Higgs sector may be necessary for 
the electroweak phase transition to be  
strongly first order.  
Therefore, we assume that the phase $\theta$ is much smaller 
than unity, so that  
sizable new $CP$-violating phenomena 
could only be induced by the phase $\alpha$.     
Then the gluino and neutralino diagrams, respectively, give dominant 
contributions to the neutron and electron EDMs,  
leading to relaxed constraints on the masses of supersymmetric 
particles \cite{edm}.  

     The $CP$-violating phase $\alpha$ is contained in the 
squark mass-squared matrices. 
Corresponding to two chiralities for the quark, there are 
two species of squark for each flavor: the left-handed squark $\sq_L$ and 
the right-handed squark $\sq_R$.  
Neglecting generation mixings, these squarks are in 
mass eigenstates in the SU(2)$\times$U(1) symmetric vacuum.  
When this symmetry is broken,  
they are mixed to form mass eigenstates $\sq_1$ and $\sq_2$, 
owing to non-vanishing 
vacuum expectation values of the Higgs bosons.   
The mass-squared matrix $M^2_q$ for the squarks corresponding to 
a quark $q$ with the mass $m_q$, the electric charge $Q_q$, 
and the third component of the weak isospin $T_{3q}$ becomes  
\[
    \lefteqn{M^2_q =} \hspace{9cm} 
\]
\[
 \left(\matrix{m_q^2 + \c2b (T_{3q} - Q_q\sw2 )M_Z^2 + \M_{qL}^2 &
                                            m_q (R_q\mH + A^*\mgr) \cr
                   m_q (R_q\mH^* + A\mgr) &
                               m_q^2 +  Q_q\c2b\sw2 M_Z^2 + \M_{qR}^2}
           \right),   
\]
\begin{eqnarray}
   R_q &= & \frac{1}{\tb}
              \quad (\ T_{3q} = \frac{1}{2}\ ),  
 \label{2} \\
       &= & \tb 
              \quad (\ T_{3q} = -\frac{1}{2}\ ), 
                       \nonumber \\
   \tb &= & \frac{\v2}{\v1},  \nonumber 
\end{eqnarray}
where $\v1$ and $\v2$ stand for the vacuum expectation values 
of the Higgs bosons;   
$\M_{qL}^2$ and $\M_{qR}^2$ 
the mass-squared parameters for $\sq_L$ and $\sq_R$; and 
$\mgr$ the gravitino mass. 
The value of $m_q$ is determined by $\v1$ for $T_{3q}=-1/2$ 
and by $\v2$ for $T_{3q}=1/2$.   
The masses for $\sq_1$ and $\sq_2$ are obtained by diagonalizing 
$M_q^2$ in Eq. (\ref{2}), which are denoted by $\M_{q1}$ and $\M_{q2}$.  
The phase $\alpha$ is also contained in the slepton mass-squared 
matrices, which are obtained by appropriately changing Eq. (\ref{2}).  

     The baryon asymmetry could be generated   
by the charge transport mechanism \cite{ctm}, if the electroweak 
phase transition of the universe is strongly first order.  
At the phase transition,  
bubbles of the broken phase nucleate in the 
SU(2)$\times$U(1) symmetric phase.  On the bubble wall,  
left-handed $t$ squarks $\st_L$ coming from the symmetric phase 
can be reflected to become right-handed $t$ squarks $\st_R$, 
and vice verse.  
Mass eigenstates of $t$ squarks $\st_1$ and $\st_2$ in the broken phase 
can be transmitted to the symmetric phase and become left- or 
right-handed $t$ squarks.  
In these processes $CP$ violation makes differences in 
reflection or transmission probabilities between $CP$ conjugate states.  
Consequently some net charges are induced in the 
symmetric phase, which otherwise should remain vanishing.   
These net charges could lead to biases on equilibrium conditions  
in front of the wall favoring a non-vanishing value for the baryon number, 
which is then realized through electroweak anomaly.   
Although other squarks or sleptons could generate  
the net charges, their contributions can be neglected 
because of their small Yukawa coupling constants.   

     The reflection and transmission rates are 
obtained by solving the Klein-Gordon equations for 
the $t$ squarks.  In the rest frame of the wall the 
equations are given by 
\begin{equation}
 \left(\matrix{
        \frac{\partial^2}{\partial t^2}+(M_t^2)_{11} & (M_t^2)_{12} \cr
         (M_t^2)_{21} &  \frac{\partial^2}{\partial t^2}+(M_t^2)_{22}
                                               }\right) 
  \left(\matrix{\st_L \cr
                         \st_R \cr}    \right) 
                   =\frac{\partial^2}{\partial z^2}   
  \left(\matrix{\st_L \cr
                         \st_R \cr}    \right), 
\label{3}
\end{equation}
where $(M_t^2)_{ij}$ represents the $(i,j)$ component 
of the $t$-squark mass-squared matrix.  
The bubble wall is taken to be parallel to the $xy$-plane and 
perpendicular to the velocity of the particles.  
In the symmetric phase the vacuum expectation values $\v1$ and 
$\v2$ vanish, while in the broken phase they are related to the  
$W$-boson mass as $M_W=(g/2)\sqrt{\v1^2+\v2^2}$.  
The vacuum expectation values vary along the $z$-axis in the wall. 
Taking the symmetric and broken phases in the regions $z<0$ and 
$2\dW<z$, respectively, we assume that the $z$-dependences 
of $\v1$ and $\v2$ are given by  
\begin{eqnarray}
   \sqrt{\v1^2+\v2^2} &=& 
           \frac{M_W}{g}\{1+\tanh(\frac{z}{\dW}-1)\pi\},   \\
\label{4}              
   \frac{\v2}{\v1} &=& (5-\frac{2z}{\dW})\tb,  
\label{5}
\end{eqnarray}
where $\tb$ represents the ratio of $\v2$ to $\v1$ in the broken phase.  
The wall width $\dW$ has been estimated in the SM as 
$\dW\sim 10/T$ \cite{turok}, 
although there are large uncertainties and model dependences.  
If the phase transition is strongly first order, the wall 
width becomes thinner.  
It is seen from Eq. (\ref{3}) that the complex phase of $(M^2_t)_{12}$ 
has to vary with $z$ in order to violate $CP$ invariance.  
This is satisfied, if the ratio $\v2/\v1$ has a $z$-dependence, 
as Eq. (\ref{5}).  The Klein-Gordon equations are solved 
numerically following the procedure given in Ref. \cite{aoki}.  

     For the charge which gives a bias on the equilibrium conditions 
in the SU(2)$\times$U(1) symmetric phase, we take hypercharge.  
Since the hypercharges of $\st_L$ and $\st_R$ 
are respectively 1/6 and 2/3, $CP$ asymmetries 
in the reflection and transmission rates lead to 
a net flux of hypercharge emitted from the bubble wall.  
In front of the wall the gauge interactions and the Yukawa 
interactions proportional to the $t$-quark mass 
are considered to be in chemical equilibrium.  
We also assume that the self interactions of Higgs bosons, 
Higgsinos, and SU(2)$\times$U(1) gauginos are 
in equilibrium, respectively.  
Among the supersymmetric particles, the gluinos are assumed 
to be heavy enough and out of equilibrium.  
Taking the densities for total baryon number,  baryon number of the 
third generation, and lepton number to be approximately vanishing, 
the chemical potential $\mu_B$ of the baryon number is related  
to the hypercharge density $\rho_Y$ through the equilibrium conditions as   
\begin{equation}
     \mu_B=-\frac{2\rho_Y}{9T^2},   
\label{6}
\end{equation}
where $T$ stands for the temperature of the 
electroweak phase transition.  
The number densities 
of the right-handed quarks and leptons except the $t$ quark 
and of their superpartners have also been taken to be vanishing.  
We can see that the non-vanishing hypercharge density 
in the symmetric phase becomes a bias for the 
baryon number density.  
    
     In the symmetric phase, baryon number can change 
through electroweak anomaly at a nonnegligible rate.  
Assuming detailed balance for the transitions among 
the states of different baryon numbers, the rate equation of the 
baryon number density $\rho_B$ is 
given by \cite{dine1} 
\begin{eqnarray}
   \frac{d\rho_B}{dt}&=&-\frac{\Gamma}{T}\mu_B,  
\label{7} \\
   \Gamma&=&3\kappa(\alpha_WT)^4 ,  \nonumber
\end{eqnarray}
where $\Gamma$ denotes the rate per unit 
time and unit volume for the transition between the neighboring 
states different by unity in baryon number, $\kappa$ being  
$0.1-1$ \cite{ambjorn}.  
From Eqs. (\ref{6}) and (\ref{7}) the baryon number density 
of the symmetric phase captured by the wall 
is roughly estimated as \cite{nelson1}  
\begin{equation} 
   \rho_B\approx\frac{2\Gamma F_Y\tau_T}{9T^3v_W}, 
\label{8}
\end{equation}
where $\tau_T$ represents the time which carriers of the hypercharge 
flux spend in the symmetric phase before captured 
by the wall; $v_W$ the velocity of the wall;  
and $F_Y$ the hypercharge flux 
emitted from the wall.  
The transport time $\tau_T$ 
may be approximated by the mean free time of the carriers. 
The rough estimate for $\tau_T$ gives values 
of order of $10/T$ for the quarks \cite{joyce},   
which would also be applicable to the squarks.  
The wall velocity $v_W$ is estimated as $v_W=0.1-1$ \cite{turok}. 
The net hypercharge flux $F_Y$ is given by 
\begin{eqnarray}
  F_Y&=&F_{\st_L}+F_{\st_R}+\sum_{i=1}^2F_{\st_i},
\label{9}    \\
  F_{\st_L}&=&-\frac{(1-\vW2)T}{(2\pi)^2}
    \int_{\M_{tL}}^\infty dE
      E\ln[1-\exp\left(
      -\frac{E-v_W\sqrt{E^2-\M_{tL}^2}}{T\sqrt{1-\vW2}}
                  \right)]            \nonumber \\
         & &\frac{2}{3}\{R(\st_L\rightarrow \st_R)
                                 -R(\st_L^*\rightarrow \st_R^*)\},  
                         \nonumber \\
 F_{\st_R}&=&-\frac{(1-\vW2)T}{(2\pi)^2}
    \int_{\M_{tR}}^\infty dE
      E\ln[1-\exp\left(
      -\frac{E-v_W\sqrt{E^2-\M_{tR}^2}}{T\sqrt{1-\vW2}}
                  \right)]            \nonumber \\
         & &\frac{1}{6}\{R(\st_R\rightarrow \st_L)
                                 -R(\st_R^*\rightarrow \st_L^*)\},  
                         \nonumber \\
 F_{\st_i}&=&-\frac{(1-\vW2)T}{(2\pi)^2}
    \int_{\M_{ti}}^\infty dE
      E\ln[1-\exp\left(
      -\frac{E+v_W\sqrt{E^2-\M_{ti}^2}}{T\sqrt{1-\vW2}}
                  \right)]            \nonumber \\
         & &[\frac{1}{6}\{R(\st_i\rightarrow \st_L)
                                 -R(\st_i^*\rightarrow \st_L^*)\}  
               +\frac{2}{3}\{R(\st_i\rightarrow \st_R)
                                 -R(\st_i^*\rightarrow \st_R^*)\}], 
                         \nonumber 
\end{eqnarray}
where $R(\st_L\rightarrow \st_R)$ etc. denote the 
probabilities for transitions at the wall.  
Although the procedure for obtaining Eq. (\ref{8})  
assumes various simplifications,  
it would still be reasonable for making only a rough estimate 
of the baryon number density.  

     The baryon number captured in the broken phase does not 
change, since there baryon number violation by electroweak 
anomaly is negligibly small.  Consequently 
the ratio of baryon number to entropy becomes constant 
afterward, which is given by 
\begin{equation}
  \frac{\rho_B}{s}=\frac{15\kappa\alpha_W^4F_Y\tau_T}
                                       {\pi^2g_*v_W T^2}, 
\label{10}
\end{equation}
where $g_*$ represents the relativistic degree of freedom 
for the particles.  For definiteness, we take $g_*=214.75$, 
where all the particles except for the gluinos are   
taken into account.  
The ratio has been observed as 
$\rho_B/s=(2-9)\times 10^{-11}$ \cite{pdg}.   

\begin{figure}
\caption{The ratio of baryon number to entropy as a function of 
$\mgr$ for $\alpha=\pi/4$ and $\theta=0$.  
 The values of $v_W$ and $\dW$ for curves (i.a)--(ii.b) 
 are given in Table \ \protect\ref{tab1}.
 The other parameters are taken for $\tb=2$, $|\mH|=100$ GeV, 
and $T=200$ GeV.  }
\label{fig1}

\vspace{2cm}
\psfig{file=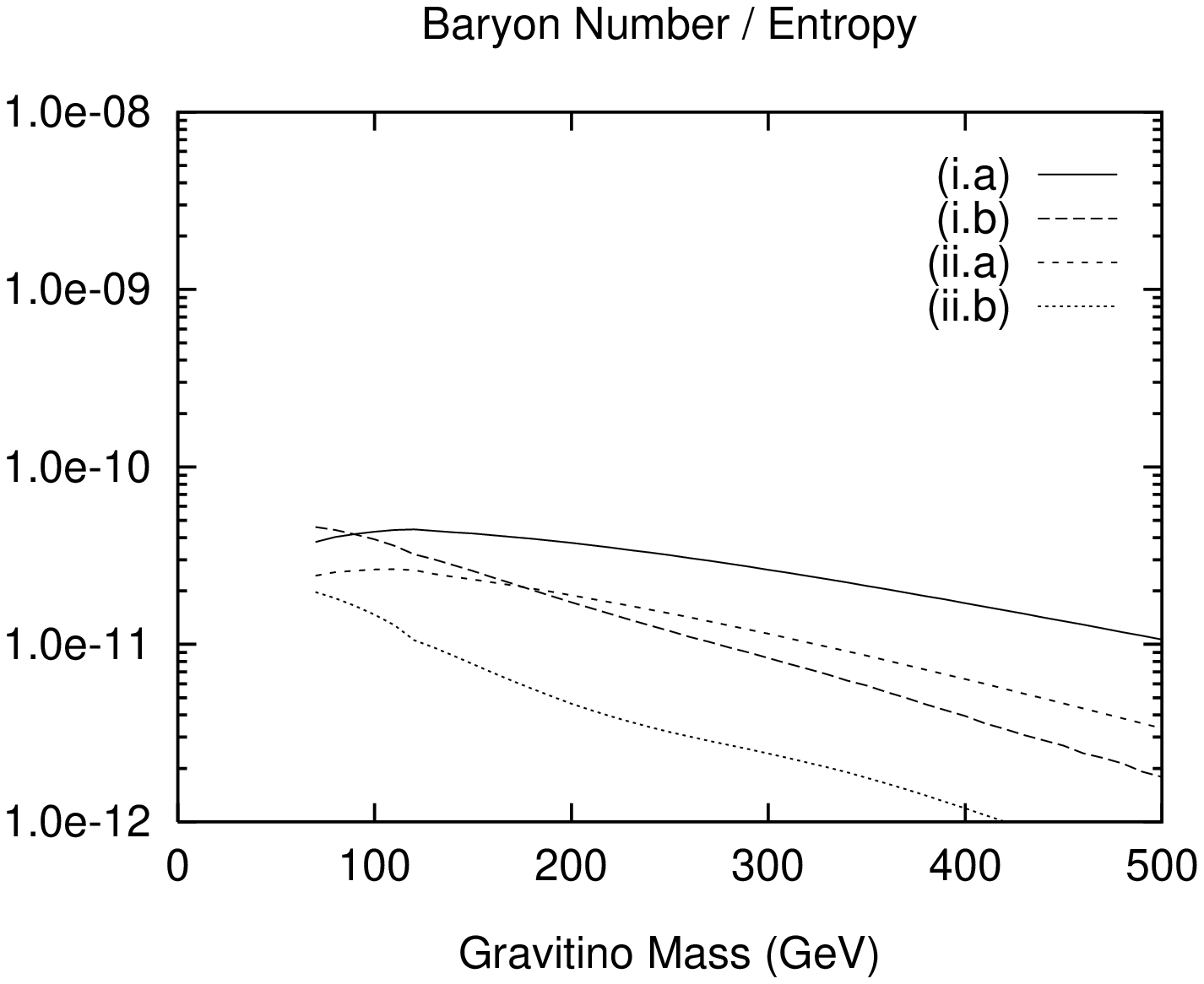}
\end{figure}

     We show the ratio $\rho_B/s$ 
in Fig. \ref{fig1} as a function of the gravitino mass  
$\mgr$ for $\alpha=\pi/4$, $\theta=0$, $\tb=2$, and 
$|\mH|=100$ GeV.   
For simplicity, we set $|A|=1$ and $\mgr=\M_{tL}=\M_{tR}$.   
In the mass ranges where curves are 
not drawn, the lightest squark has a mass 
smaller than 45 GeV, which is ruled out by 
experiments at LEP \cite{pdg}.  
The temperature is taken for $T=200$ GeV.   
We take four sets of 
values for $v_W$ and $\dW$ listed in Table \ref{tab1},  
which correspond to four curves (i.a)$-$(ii.b).  
For definiteness we set $\tau_T=10/T$ and $\kappa=1$.  
The resultant ratio can be compatible with the observed value,  
if the wall width is thin and $\kappa$ is of order unity.   
The gravitino mass should be of order 100 GeV, whereas   
the ratio does not vary much with $|\mH|$ and $\tb$.  
If the $CP$-violating phase $\alpha$ is  
smaller than $0.1$, it is difficult to generate 
an enough amount of asymmetry through this mechanism.  
The sign of $\rho_B$ depends on the shape of $\v2/\v1$ 
in the wall as well as $\alpha$.   

     In the parameter ranges where the baryon asymmetry can 
be explained, the EDMs of the neutron and the electron receive 
large contributions, respectively, 
from the gluino and the neutralino diagrams.   
The gluino contribution to the quark EDM is given by \cite{edm} 
\begin{eqnarray}
     d_q^G/e &=& \frac{2\alpha_S}{3\pi}\left(\sin\alpha|A|
          - R_q\sin\theta\frac{|\mH |}{\mgr}\right)    
   \frac{\mgr m_q}{\M_q^3}Q_q\frac{\mg}{\M_q}
             K\left(\frac{\mg^2}{\M_q^2}\right), 
 \label{11} \\
K(r) &=& \frac{-1}{2(1 - r)^3}[1 + 5r +
                             \frac{2r(2 + r)}{1 - r}\ln r],   
                 \nonumber 
\end{eqnarray}
where $\mg$ and $\M_q$ denote the gluino mass and 
the average mass for $\M_{q1}$ and $\M_{q2}$, respectively.  
Assuming the same mass for the gauginos at the grand unification 
scale, $\mg$ is related to the SU(2) gaugino mass $\m2$ by 
$\mg=\sw2(\alpha_S/\alpha_{EM})\m2$.  
Since $\theta$ is taken for zero, the chargino contribution 
to the EDM is negligible.  
From the non-relativistic quark model, the EDM of the 
neutron $d_n$ is given by $d_n=(4d_d-d_u)/3$, 
where $d_u$ and $d_d$ are respectively the EDMs of the 
$u$ quark and the $d$ quark.    

\begin{figure}
\caption{The electric dipole moment of the neutron
   as a function of $\mgr$ for $\alpha=\pi/4$ and $\theta=0$.  
   Three curves correspond to three values for $\m2$: 
(i) 200 GeV, (ii) 500 GeV, (iii) 1 TeV.  
 The other parameters are taken for $\tb=2$ and $|\mH|=100$ GeV. }
\label{fig2}

\vspace{2cm}
\psfig{file=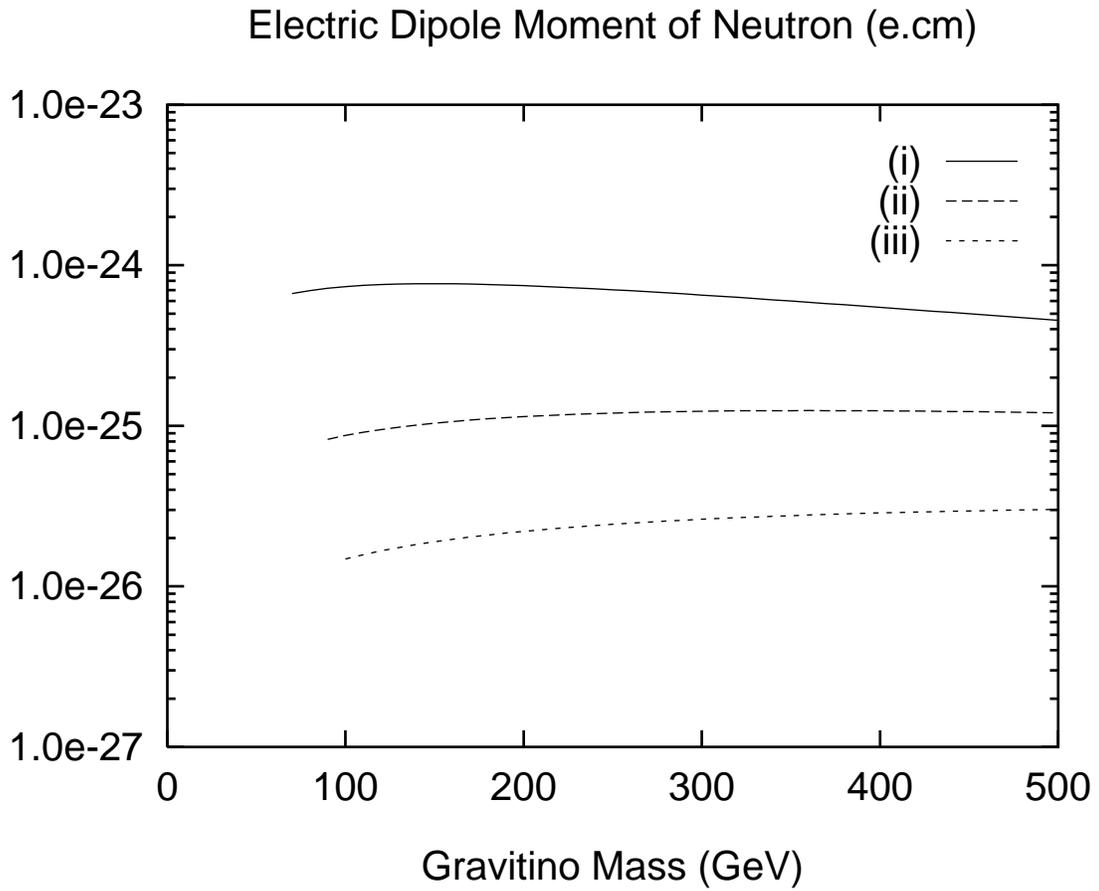}
\end{figure}

     In Fig. \ref{fig2} 
the neutron EDM is shown as a function of $\mgr$ 
for the same values of $\alpha$, $\theta$, $\tb$, and $|\mH|$ 
as those in Fig. \ref{fig1}.   
We also set $|A|=1$ and $\mgr=\M_{qL}=\M_{qR}$ 
in the mass-squared matrices for the $u$ squarks and the $d$ squarks.    
Three curves correspond to  
three values for $\m2$: (i)200 GeV, (ii)500 GeV, (iii)1 TeV.  
In the mass ranges where curves are not drawn, 
the lightest squark is lighter than either 45 GeV or 
the lightest neutralino, the latter of which is disfavored by cosmology.    
The experimental upper bound on the magnitude of the neutron EDM 
is about $1\times 10^{-25} e$cm \cite{pdg}.  
For $\m2=500-1000$ GeV the EDM has a value of 
$10^{-25}-10^{-26} e$cm, which is consistent with the experimental 
constraint though not much smaller than the upper bound.  
For the same parameter ranges the magnitude of 
the electron EDM is also not much 
smaller than its present experimental upper bound.  

     We have discussed whether baryon asymmetry of the 
universe could be generated at the electroweak phase transition 
in the SSM.  If one $CP$-violating phase $\alpha$ is not suppressed 
while another phase $\theta$ being kept small, the $t$ squarks 
with the masses of order 100 GeV efficiently mediate 
the charge transport mechanism.  
An enough amount of asymmetry could 
be induced, provided that the bubble wall is thin 
and the value of $\kappa$ for the baryon number violation rate 
is not much smaller than unity.  
The masses of the other squarks and sleptons are of the 
the same order of magnitude as the $t$-squark masses.   
For $\m2$, $|\mH|\lsim 1$ TeV the charginos and neutralinos have masses 
of $100-1000$ GeV, whereas the gluinos are heavier than 1 TeV.  
For these parameter values the electroweak 
phase transition of the universe may be strongly first order \cite{ewpt},  
which is necessary for the charge transport mechanism.  
If the $CP$-violating phase $\theta$ is not suppressed, 
the charge transport mechanism can be mediated by the charginos with 
the masses of order 100 GeV instead of the $t$ squarks, providing 
an enough amount of asymmetry \cite{aoki}.  
In this case the squarks and sleptons have masses 
larger than 1 TeV, which may necessitate  
some extension for the Higgs sector of the SSM in order for   
the phase transition to be strongly first order.   

     The charge transport mechanism for baryogenesis in the SSM 
implies that the neutron and electron EDMs have 
magnitudes not much smaller than their present 
experimental upper bounds.  For $\alpha\sim 1$, $\theta\ll 1$ 
the gluino diagrams give dominant contributions to the 
EDM of the neutron, while for $\theta\sim 1$ do the chargino diagrams.    
In both cases, the observed baryon asymmetry leads to the prediction 
for the neutron EDM of $10^{-25}-10^{-26} e$cm. 
The EDM of the electron also receives a large contribution  
from the neutralino diagram for $\alpha\sim 1$, $\theta\ll 1$ 
or from the chargino diagram for $\theta\sim 1$.  
These predicted values would be 
within reach of experiments in the near future.  
In particular, the superthermal method for the production 
of ultracold neutrons \cite{yoshiki} could improve 
the accuracy of measurements for the neutron EDM  
by one order of magnitude.  
Possible detection of the EDMs would make the SSM a more 
credible candidate for the theory at the electroweak scale.  

\section*{Acknowledgments}

     We thank J. Arafune and K. Yamamoto for many valuable discussions.  
The work of M.A. is supported in part by the Grant-in-Aid for Scientific 
Research from the Ministry of Education, Science and Culture, Japan.  
This work is supported in part by 
the Grant-in-Aid for Scientific Research (No. 08640357) and by  
the Grant-in-Aid for Joint Scientific Research (No. 08044089) 
from the Ministry of Education, Science and Culture, Japan.

\newpage 
\vspace{2cm}

\begin{table}
\caption{The values of $v_W$ and $\dW$ for curves 
            (i.a)--(ii.b) in Fig.\ \protect\ref{fig1}.} 
\label{tab1}

\vspace{1cm}
\begin{center}
\begin{tabular}{ccccc}
\hline
\hline
    & (i.a) & (i.b) & (ii.a) &(ii.b) \\
\hline
 $v_W$  &  0.1 & 0.1 & 0.6 & 0.6 \\
 $\dW$  & 1/T  & 5/T  & 1/T & 5/T  \\
\hline
\hline
\end{tabular} 
\end{center}
\end{table}
\pagebreak

\vspace{2cm}

\end{document}